\documentclass{article}
\usepackage[portuguese]{babel}
\usepackage[latin1]{inputenc}
\usepackage{amssymb}
\usepackage{amsmath}
\usepackage{fancyhdr}
\usepackage{verbatim}
\usepackage{lastpage}
\usepackage{ifthen}
\usepackage[pdftex]{color,graphicx}

\usepackage{arxiv}

\usepackage[latin1]{inputenc}
\usepackage{url}            

\usepackage{soul}
\usepackage[normalem]{ulem}
\newcommand\redout{\bgroup\markoverwith
{\textcolor{red}{\rule[.5ex]{2pt}{0.4pt}}}\ULon}
\usepackage{xcolor}
\usepackage[normalem]{ulem}
\usepackage{caption}
\usepackage{graphicx}
\usepackage{color,soul}
\usepackage{subfig}

\title{User Questions from Tweets on COVID-19: An Exploratory Study}

\author{
  Tiago de Melo \\
  Laborat\'{o}rio de Sistemas Inteligentes (LSI)\\
  Universidade do Estado do Amazonas (UEA)\\
  Amazonas, Brazil \\
  \texttt{tmelo@uea.edu.br} 
}

\begin{document}
\maketitle

\begin{abstract}
Social media platforms, such as Twitter, provide a suitable avenue for users (people or patients) concerned on health questions to discuss and share information with each other. In December 2019, a few coronavirus disease cases were first reported in China. Soon after, the World Health Organization (WHO) declared a state of emergency due to the rapid spread of the virus in other parts of the world. In this work, we used automated extraction of COVID-19 discussion from Twitter and a natural language processing (NLP) method based on topic modeling to discover the main questions related to COVID-19 from tweets. Moreover, we created a Named Entity Recognition (NER) model to identify the main entities of four different categories: disease, drug, person, and organization. Our findings can help policy makers and health care organizations to understand the issues of people on COVID-19 and it can be used to address them appropriately.
\end{abstract}


\section{Introdu\c{c}\~{a}o}

Em dezembro de 2019, o surto do COVID-19 na China foi noticiado~\cite{malta2020coronavirus}. Devido ao r\'{a}pido espalhamento do v\'{i}rus no mundo, a Organiza\c{c}\~{a}o Mundial de Sa\'{u}de (OMS) declarou estado de emerg\^{e}ncia. Recentes pesquisas confirmaram que a atual pandemia pode dobrar o n\'{u}mero de pessoas infectadas a cada 7 (sete) dias e que cada paciente pode espalhar o v\'{i}rus, na m\'{e}dia, para 2.2 outras pessoas~\cite{velavan2020covid}. Na Am\'{e}rica Latina, houve o registro de 718,615 casos de COVID-19 e 39,229 mortes confirmadas at\'{e} o dia 25 de maio de 2020~\cite{world2020coronavirus}. No continente, o Brasil \'{e} o pa\'{i}s mais afetado pela doen\c{c}a. De acordo com o mesmo relat\'{o}rio~\cite{world2020coronavirus}, houve o registro de 347,398 casos de infectados e de 22,013 mortes no Brasil.

Devido a propaga\c{c}\~{a}o da doen\c{c}a no mundo, as plataformas de m\'{i}dias sociais como Twitter, Facebook e Instagram tornaram-se locais onde ocorre uma intensa e cont\'{i}nua troca de informa\c{c}\~{o}es entre \'{o}rg\~{a}os governamentais, profissionais da \'{a}rea de sa\'{u}de e o p\'{u}blico em geral. Um representativo n\'{u}mero de estudos cient\'{i}ficos t\^{e}m mostrado que as m\'{i}dias sociais podem desempenhar um papel importante como fonte de dados para an\'{a}lise de crises e tamb\'{e}m para entender atitudes e comportamentos das pessoas durante uma pandemia~\cite{li2020characterizing,kim2016topic,du2019twitter}.

Com o objetivo de auxiliar o monitoramento da sa\'{u}de p\'{u}blica e tamb\'{e}m para dar suporte a tomada de decis\~{a}o de profissionais, diversos sistemas de monitoramento v\^{e}m sendo desenvolvidos para classificar grandes quantidades de dados oriundos das m\'{i}dias sociais. Estes dados podem ser empregados para identificar rapidamente os pensamentos, atitudes, sentimentos e t\'{o}picos que ocupam as mentes das pessoas em rela\c{c}\~{a}o \`{a} pandemia do COVID-19~\cite{abd2020top}. A an\'{a}lise sistem\'{a}tica desses dados pode ajudar os governantes, profissionais da sa\'{u}de e o p\'{u}blico em geral a identificar quest\~{o}es que mais lhes interessam e trat\'{a}-las de maneira mais apropriada.

Dentre as plataformas de m\'{i}dias sociais, o Twitter \'{e} uma das mais populares. De acordo com~\cite{gopi2020classification}, existe aproximadamente 200 milh\~{o}es de usu\'{a}rios registrados nesta plataforma e que publicam mais de 500 milh\~{o}es de tu\'{i}tes diariamente. Portanto, pode-se aproveitar desse alto volume e troca frequente de informa\c{c}\~{o}es para se conhecer as d\'{u}vidas sobre determinadas crises. Como exemplo de import\^{a}ncia desta plataforma em situa\c{c}\~{o}es de crise, a atual pandemia de COVID-19 foi primeiro comunicada para a populac\~{a}o na China atrav\'{e}s do site Weibo, que \'{e} o correspondente ao Twitter na China, antes mesmo do pronunciamento oficial das autoridades locais. Recentemente, existe um grande interesse de pesquisadores investigando o uso do Twitter para pesquisas relacionadas \`{a} sa\'{u}de p\'{u}blica~\cite{abd2020top,li2020characterizing,ordun2020exploratory,liu2020health}.

Diante deste cen\'{a}rio, n\'{o}s apresentamos um estudo explorat\'{o}rio de minera\c{c}\~{a}o de opini\~{a}o das mensagens de usu\'{a}rios do Twitter relacionadas \`{a} COVID-19. Mais especificamente, n\'{o}s focamos a nossa an\'{a}lise nas perguntas dos usu\'{a}rios, pois entendemos que seja um tipo de mensagem apropriado para se compreender as principais d\'{u}vidas das pessoas sobre a atual pandemia. A nossa an\'{a}lise se concentrar\'{a} em postagens em portugu\^{e}s pelo fato do Brasil ser o pa\'{i}s mais populoso da Am\'{e}rica Latina e tamb\'{e}m o pa\'{i}s mais afetado pela COVID-19. Para isto, n\'{o}s coletamos e processamos os tu\'{i}tes postados em portugu\^{e}s de 1$^o$ de janeiro a 30 de abril de 2020. Os tu\'{i}tes coletados foram processados e as perguntas foram identificadas. N\'{o}s analisamos os tu\'{i}tes coletados usando a t\'{e}cnica de modelagem de t\'{o}picos para identificar os principais t\'{o}picos discutidos pelas pessoas no Twitter. N\'{o}s ainda desenvolvemos um modelo de Reconhecimento de Entidades Mencionadas (REM) que permite identificar as principais men\c{c}\~{o}es a um grupo pr\'{e}-definido de entidades: a) doen\c{c}as; b) medicamentos; c) pessoas; d) organiza\c{c}\~{o}es. A an\'{a}lise desses dados pode ajudar os formuladores de pol\'{i}ticas p\'{u}blicas e as organiza\c{c}\~{o}es de assist\^{e}ncia m\'{e}dica a compreender as principais quest\~{o}es do p\'{u}blico em geral.

Dentre as nossas descobertas, n\'{o}s conseguimos identificar a mudan\c{c}a da percep\c{c}\~{a}o dos usu\'{a}rios, ao longo do tempo, em rela\c{c}\~{a}o \`{a} doen\c{c}a. A preocupa\c{c}\~{a}o com a morte somente ficou evidenciada ap\'{o}s o registro, em mar\c{c}o de 2020, do primeiro caso de brasileiro morto por COVID-19. Foi poss\'{i}vel ainda perceber que, ao aplicarmos a identifica\c{c}\~{a}o das entidades nomeadas das perguntas, h\'{a} muita d\'{u}vida sobre os tipos de medicamentos que poderiam ser utilizados para combater a doen\c{c}a. A identifica\c{c}\~{a}o das d\'{u}vidas mais comuns sobre o uso dos medicamentos poderia ajudar os agentes p\'{u}blicos no enfretamento da doen\c{c}a atrav\'{e}s, por exemplo, de campanhas publicit\'{a}rias para elucida\c{c}\~{a}o dos principais questionamentos.

Resumidamente, as contribui\c{c}\~{o}es deste trabalho s\~{a}o:
\begin{itemize}
    \item Estudo sobre o uso de modelagem de t\'{o}picos baseado em LDA para descoberta de t\'{o}picos relativos \`{a} COVID-19 nas quest\~{o}es postadas por usu\'{a}rios do Twitter.
    \item O desenvolvimento de um modelo efetivo de reconhecimento de entidades mencionadas para o dom\'{i}nio da COVID-19.
    \item Um conjunto de postagens relacionadas \`{a} COVID-19 em portugu\^{e}s. Esses dados podem ser utilizados por pesquisadores para avaliar o sentimento das pessoas sobre a pandemia e tamb\'{e}m para outras tarefas relacionadas \`{a} sa\'{u}de p\'{u}blica. O conjunto de dados est\'{a} dispon\'{i}vel em \url{https://data.mendeley.com/drafts/sch72cpyjv}.
\end{itemize}

O trabalho est\'{a} organizado da seguinte maneira. Na Se\c{c}\~{a}o~\ref{sec:related-work} s\~{a}o apresentados trabalhos relacionados ao tema proposto. Na Se\c{c}\~{a}o~\ref{sec:metodo} \'{e} apresentada a metodologia desenvolvida neste trabalho que inclui a descri\c{c}\~{a}o da base de dados e os m\'{e}todos empregados. Na Se\c{c}\~{a}o~\ref{sec:resultados} s\~{a}o apresentados os resultados obtidos nos experimentos e uma discuss\~{a}o sobre esses resultados. Por fim, na Se\c{c}\~{a}o~\ref{sec:conclusoes} s\~{a}o apresentadas as conclus\~{o}es e trabalhos futuros.

\section{Trabalhos Relacionados}
\label{sec:related-work}

Uma mir\'{i}ade de estudos v\^{e}m investigando as quest\~{o}es publicadas por usu\'{a}rios em redes sociais. Zhao e Mei~\cite{zhao2013questions} apresentaram um estudo que analisou as necessidades publicadas por usu\'{a}rios em redes sociais. Para isto, os autores analisaram um representativo volume de quest\~{o}es postadas pelos usu\'{a}rios do Twitter. J\'{a} Paul \textit{et al.}~\cite{paul2011twitter} conduziram um estudo sobre os tipos de perguntas que os usu\'{a}rios postam no Twitter. Enquanto que Soulier \textit{et al.}~\cite{soulier2016answering} desenvolveram um m\'{e}todo para gerar respostas autom\'{a}ticas a perguntas publicadas no Twitter. 

Mais recentemente, diversos autores v\^{e}m considerando as mensagens publicadas no Twitter como fonte de dados para lidar com graves problemas sociais, como desastres e pandemias. Em~\cite{zahra2017geographic}, os autores investigaram o uso das caracter\'{i}sticas dos usu\'{a}rios do Twitter baseadas em diferentes localiza\c{c}\~{o}es durante desastres. Eles examinaram a atividade dos usu\'{a}rios desta plataforma durante os terremotos na It\'{a}lia e em Myanmar. Sinnenberg \textit{et al.}~\cite{sinnenberg2017twitter} desenvolveram um estudo sobre o uso do Twitter na sa\'{u}de p\'{u}blica. Para isso, os autores definiram uma taxonomia para descrever o uso do Twitter e caracterizar o estado atual da plataforma na pesquisa de sa\'{u}de p\'{u}blica. Mais recentemente, Alaa \textit{et al.}~\cite{abd2020top} desenvolveram um estudo cujo objetivo foi identificar os principais t\'{o}picos relacionados \`{a} pandemia de COVID-19 entre as postagens dos usu\'{a}rios do Twitter. Os autores tamb\'{e}m fizeram uso do algoritmo de \textit{Latent Dirichlet Allocation} (LDA). 

Apesar dos trabalhos relacionados assumirem a import\^{a}ncia de considerar as m\'{i}dias sociais, nenhum destes trabalhos foca em perguntas sobre a pandemia de COVID-19. Na verdade, para o melhor de nosso conhecimento, esse \'{e} o primeiro trabalho que analisou um tipo espec\'{i}fico de postagem, qual seja, as perguntas em portugu\^{e}s do Brasil.

\section{Materiais e M\'{e}todos}
\label{sec:metodo}

\subsection{Coleta de Dados}

N\'{o}s utilizamos a biblioteca Twitterscraper~\cite{twitterscraper:2020} de Python para coletar os tu\'{i}tes relacionados \`{a} COVID-19 em portugu\^{e}s no per\'{i}odo de 1$^o$ de janeiro a 30 de abril de 2020. Com o objetivo de coletar as mensagens que estavam em portugu\^{e}s, n\'{o}s utilizamos a op\c{c}\~{a}o \texttt{--lang} do Twitterscraper. N\'{o}s utilizamos um conjunto de palavras-chaves para identificar os tu\'{i}tes que faziam men\c{c}\~{a}o \`{a} COVID-19. Mais especificamente, foram consideradas as seguintes palavras-chaves: corona, coronav\'{i}rus, COVID, COVID19, COVID-19, distanciamento social, isolamento, \textit{lockdown}, quarentena, cloroquina, hidroxicloroquina, ivermectina, tamiflu, azitromicina, pandemia e comorbidade. A escolha das palavras-chaves foi baseada na lista dos termos relacionados \`{a} COVID mais frequentemente usados pelas pessoas na busca por textos publicados na Web~\cite{googletrends:2020}. 

\begin{figure}[!h]
\centering
\includegraphics[width=0.5\textwidth]{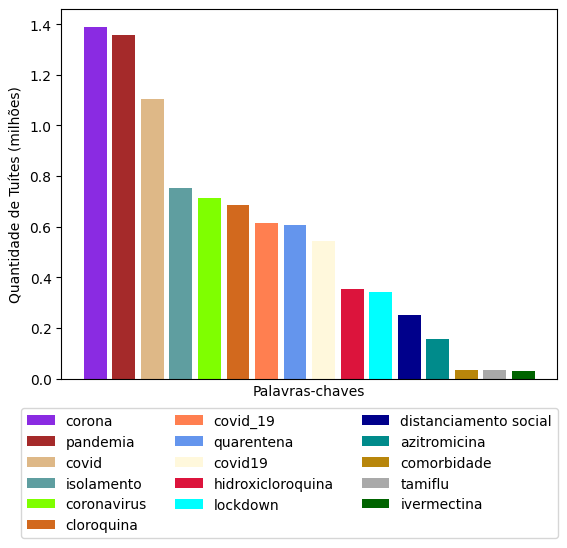}
\caption{Distribui\c{c}\~{a}o do n\'{u}mero de postagens agrupadas por palavra-chave (\textit{keyword}) durante o per\'{i}odo de janeiro a abril de 2020.}
\label{fig:distribuicao-keywords}
\end{figure}

Ap\'{o}s os tu\'{i}tes terem sido coletados, n\'{o}s usamos o Google Translator para identificar o idioma das mensagens e garantir que as mensagens foram escritas em portugu\^{e}s. Os tu\'{i}tes que n\~{a}o estavam em portugu\^{e}s foram removidos. Ap\'{o}s essa etapa, a nossa cole\c{c}\~{a}o de dados apresentou 2,619,215 tu\'{i}tes. A Figura~\ref{fig:distribuicao-keywords} mostra a distribui\c{c}\~{a}o das mensagens coletadas para cada palavra-chave. Neste gr\'{a}fico, \'{e} poss\'{i}vel observar que \textit{corona} e \textit{COVID} s\~{a}o as formas mais comuns de nomear a doen\c{c}a. Al\'{e}m disso, \'{e} poss\'{i}vel observar que cloroquina \'{e} o medicamento mais popular entre os usu\'{a}rios.


\begin{figure}[!h]
\centering
\includegraphics[width=0.5\textwidth]{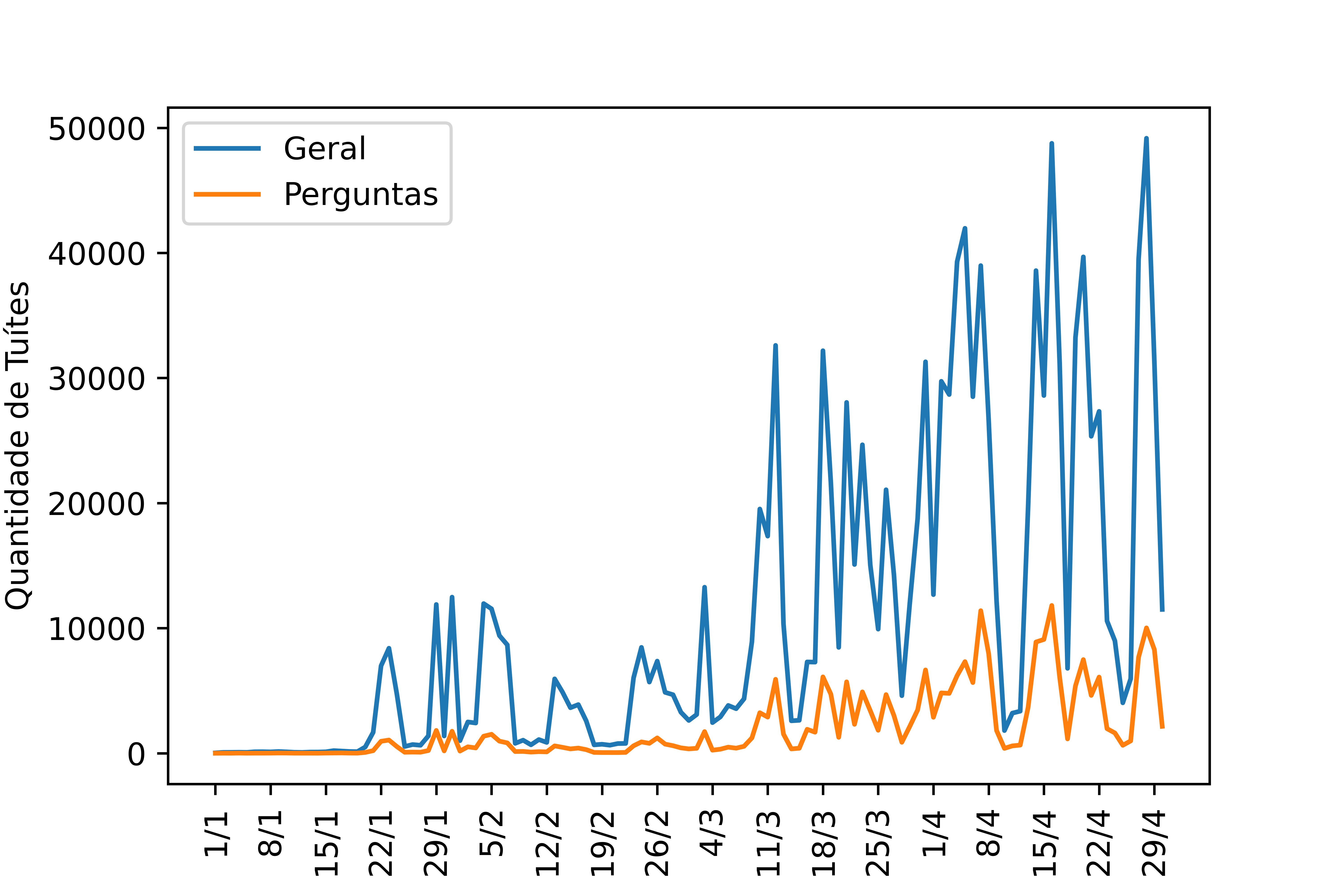}
\caption{Distribui\c{c}\~{a}o do n\'{u}mero de postagens durante o per\'{i}odo de janeiro a abril de 2020.}
\label{fig:distribuicao-tuites}
\end{figure}

Com o objetivo de identificar as perguntas dos usu\'{a}rios nas mensagens, n\'{o}s segmentamos os textos em senten\c{c}as e aplicamos um conjunto de express\~{o}es regulares implementadas em Python. As senten\c{c}as que terminavam com o caracter ? foram consideradas uma pergunta. Li \textit{et al.}~\cite{li2011question} fizeram uma avalia\c{c}\~{a}o experimental e concluiu que o m\'{e}todo de identifica\c{c}\~{a}o de quest\~{o}es baseada em regras apresentou resultados superiores ao compararmos com a abordagem baseada em aprendizagem de m\'{a}quina. Ap\'{o}s essa etapa, a nossa cole\c{c}\~{a}o de dados apresentou 2,313,070 perguntas. Figura~\ref{fig:distribuicao-tuites} mostra a distribui\c{c}\~{a}o di\'{a}ria de tu\'{i}tes coletados. A linha azul mostra a distribui\c{c}\~{a}o das postagens sobre a COVID-19, enquanto que a linha vermelha mostra a distribui\c{c}\~{a}o do n\'{u}mero de perguntas naquelas postagens. De acodo com o gr\'{a}fico, a quantidade de perguntas segue o fluxo da quantidade geral de postagens sobre a doen\c{c}a. \'{E} poss\'{i}vel ainda notar que houve um crescimento sobre a discuss\~{a}o relativa \`{a} COVID-19 a partir de meados do m\^{e}s de mar\c{c}o. Isto ocorreu ap\'{o}s o registro do primeiro brasileiro morto por causa da doen\c{c}a no dia 17 de mar\c{c}o de 2020~\cite{BBC:2020}. O gr\'{a}fico ainda mostra que as pessoas postam menos nos finais de semana ou feriado. Estes dias s\~{a}o identificados como os pontos mais baixos do gr\'{a}fico.

\subsection{Pr\'{e}-Processamento dos Dados}

O pr\'{e}-processamento dos dados \'{e} uma etapa necess\'{a}ria para uso das mensagens do Twitter. Inicialmente, n\'{o}s fizemos uma limpeza dos textos que inclui a remoc\~{a}o de \textit{hyperlinks}, e-mails, \textit{hashtags}, caracteres que indicam a replicac\~{a}o de tu\'{i}tes (RT) e duplos espa\c{c}os em branco. Ent\~{a}o n\'{o}s removemos a pontua\c{c}\~{a}o desnecess\'{a}ria e convertemos cada pergunta em uma lista de palavras. Posteriormente, n\'{o}s removemos os termos que n\~{a}o possuem relev\^{a}ncia sem\^{a}ntica (\textit{stopwords}). Em seguida, n\'{o}s usamos o pacote Gensim~\cite{rehurek_lrec} para a gera\c{c}\~{a}o de bigramas e trigramas. Finalmente, n\'{o}s aplicamos um m\'{e}todo de lematizac\~{a}o cujo o objetivo foi de reduzir as palavras a seu radical, quando poss\'{i}vel. Dentre os termos gerados, n\'{o}s mantivemos apenas os termos classificados como substantivos, adjetivos, verbos e adv\'{e}rbios.

\subsection{Modelagem de T\'{o}picos}

Modelagem de t\'{o}picos \'{e} uma das t\'{e}cnicas mais empregadas na minera\c{c}\~{a}o de dados, descoberta de dados latentes, e identifica\c{c}\~{a}o de relacionamentos entre dados e documentos textuais~\cite{jelodar2019latent}. N\'{o}s adotamos o modelo \textit{Latent Dirichlet Allocation} (LDA)~\cite{blei2003latent} para identificar os t\'{o}picos relevantes sobre a COVID-19 no nosso conjunto de dados. LDA permite descobrir t\'{o}picos latentes usando a distribui\c{c}\~{a}o de probabilidade multinomial dos termos em documentos n\~{a}o estruturados. Similarmente aos m\'{e}todos descritos em~\cite{syed2017full} e~\cite{ordun2020exploratory}, n\'{o}s executamos os experimentos variando o n\'{u}mero de t\'{o}picos. Nos nossos experimentos, n\'{o}s aplicamos a varia\c{c}\~{a}o de t\'{o}picos de 1 a 60, e selecionamos o modelo com o maior valor de pontua\c{c}\~{a}o de coer\^{e}ncia (\textit{coherence score}). A gera\c{c}\~{a}o dos t\'{o}picos foi executada para cada m\^{e}s com o objetivo de identificar a mudan\c{c}a dos t\'{o}picos discutidos no Twitter ao longo do quadrimestre. 

\begin{figure}[!h]
\centering
\includegraphics[width=0.5\textwidth]{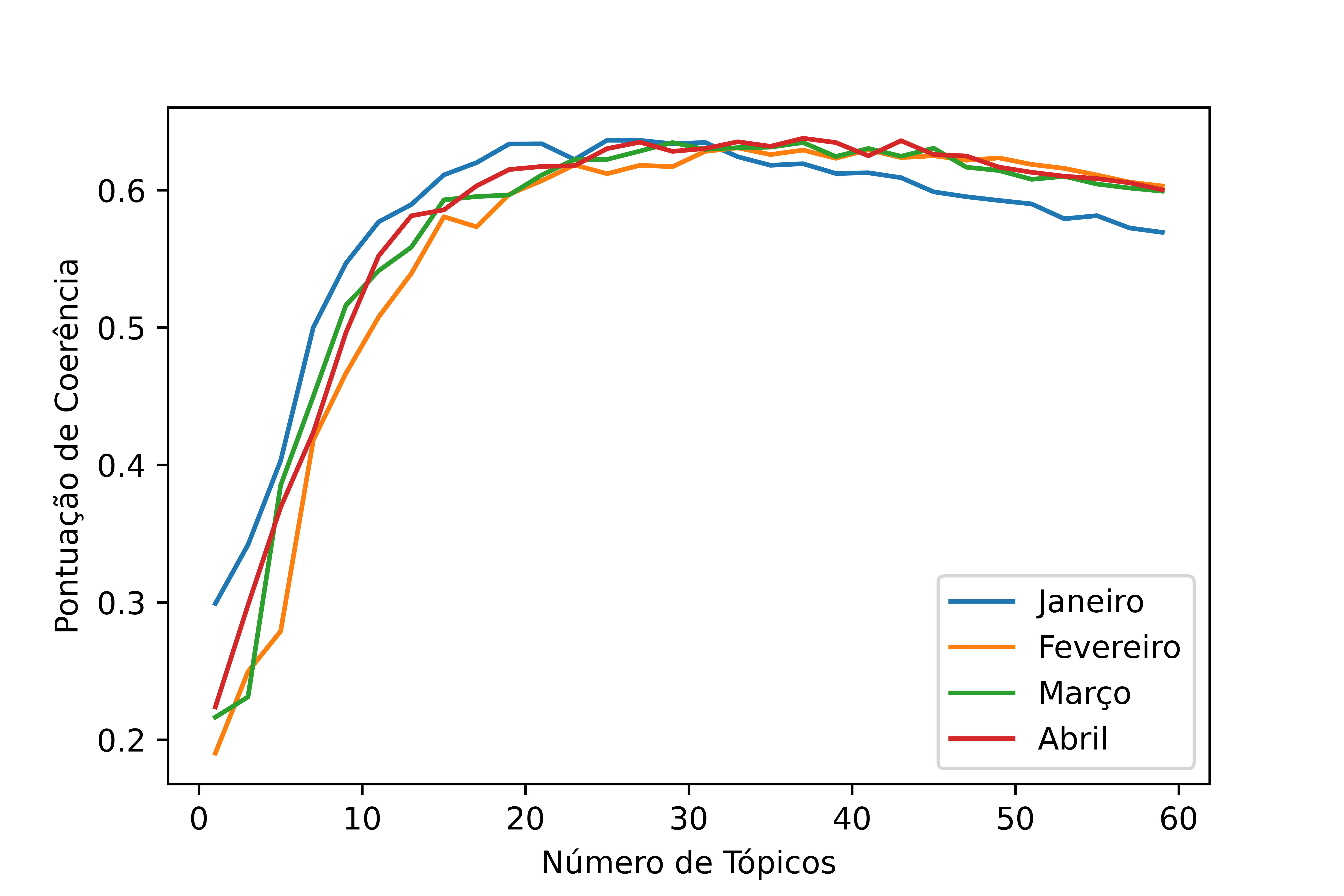}
\caption{Pontua\c{c}\~{a}o de coer\^{e}ncia pelo n\'{u}mero de t\'{o}picos para cada m\^{e}s.}
\label{fig:coerencia}
\end{figure}

N\'{o}s selecionamos o modelo que gerou 20 t\'{o}picos, com um valor m\'{e}dio de pontua\c{c}\~{a}o de coer\^{e}ncia entre os quatro meses de 0.674. Este valor \'{e} usado como uma m\'{e}trica que calcula a concord\^{a}ncia de um conjunto de pares e subconjunto de palavras e as probabilidades associadas em um \'{u}nico valor~\cite{roder2015exploring}. Em geral, os t\'{o}picos s\~{a}o interpretados como sendo coerentes se todos os termos, ou a maioria destes, s\~{a}o relacionados. A Figura~\ref{fig:coerencia} apresenta o gr\'{a}fico da rela\c{c}\~{a}o do n\'{u}mero de t\'{o}picos com a pontua\c{c}\~{a}o de coer\^{e}ncia. \'{E} poss\'{i}vel notar que a pontua\c{c}\~{a}o de coer\^{e}ncia tende a reduzir quando o n\'{u}mero de t\'{o}picos seja superior a 20.

\begin{figure}[!h]
    \centering
  \subfloat[Janeiro\label{1a}]{%
       \includegraphics[width=0.5\linewidth]{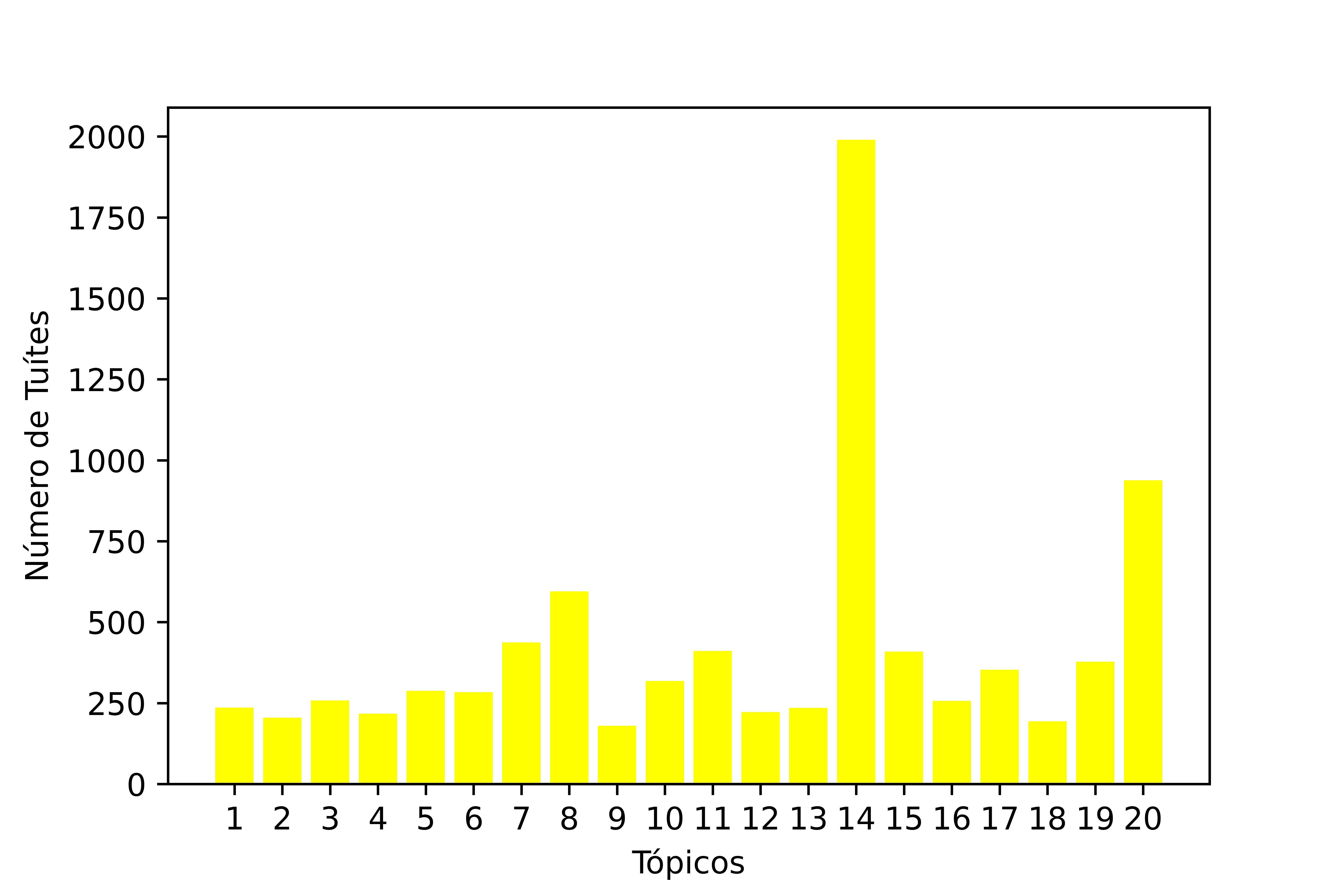}}
    \hfill
  \subfloat[Fevereiro\label{1b}]{%
        \includegraphics[width=0.5\linewidth]{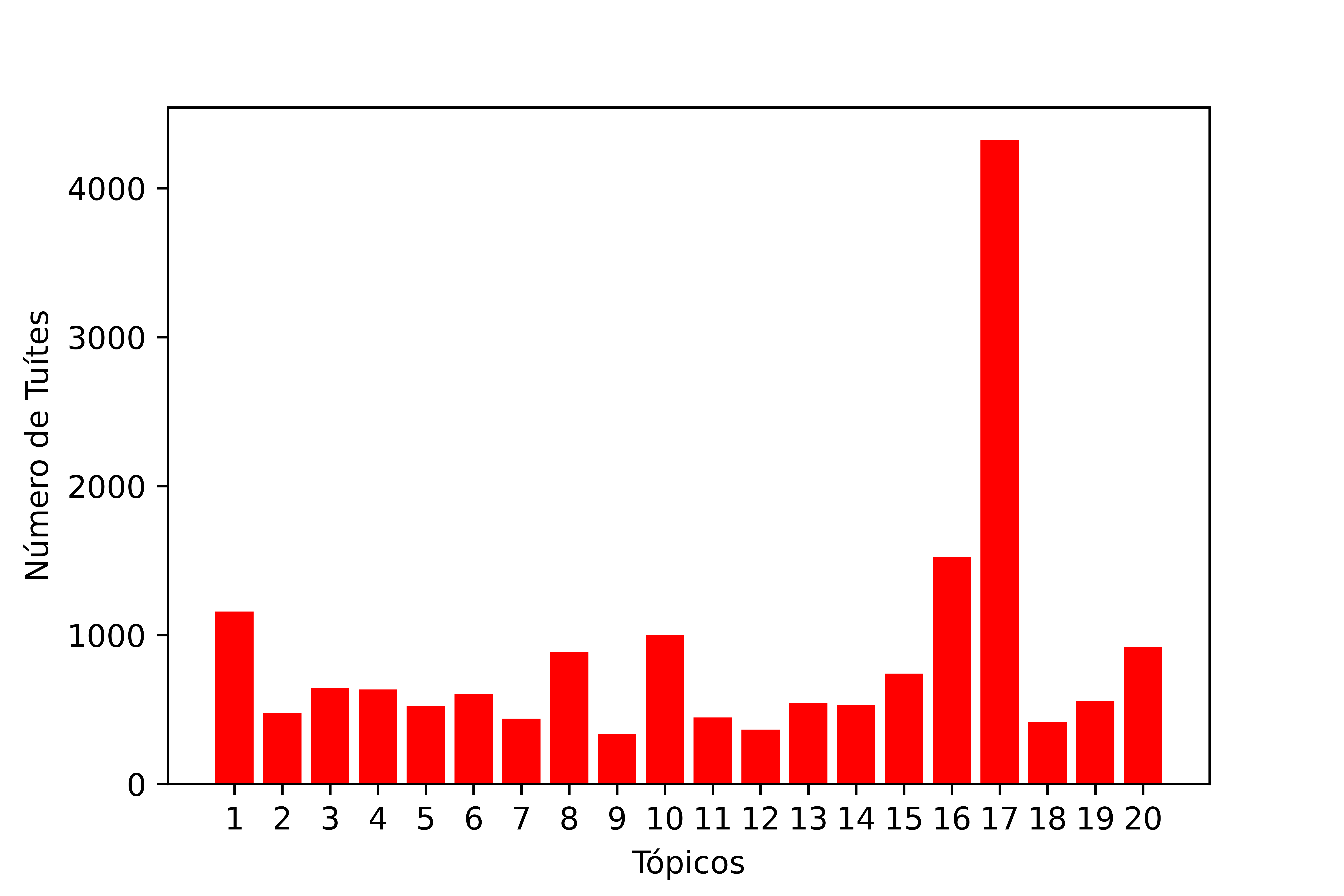}}
    \\
  \subfloat[Mar\c{c}o\label{1c}]{%
        \includegraphics[width=0.5\linewidth]{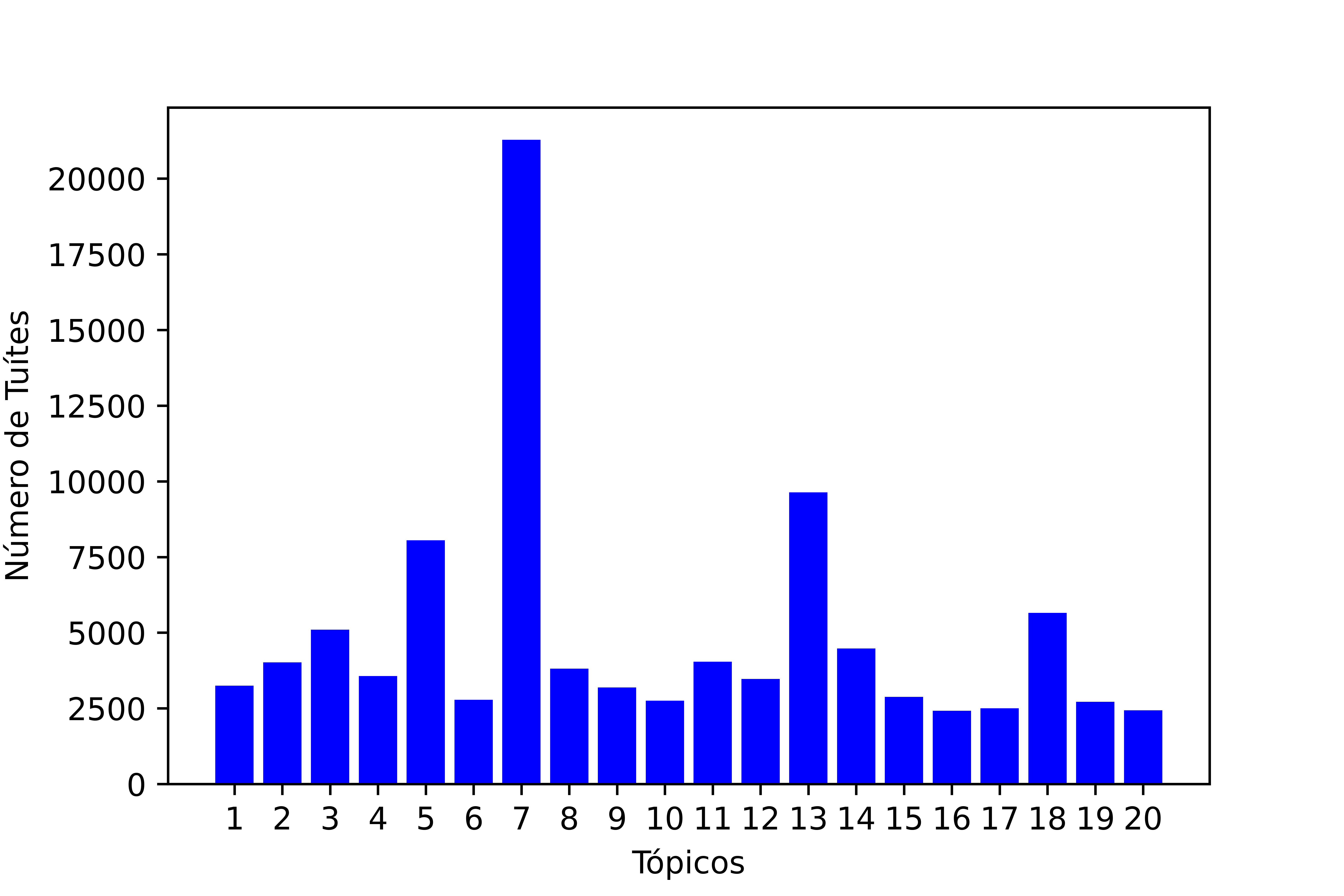}}
    \hfill 
  \subfloat[Abril\label{1d}]{%
        \includegraphics[width=0.5\linewidth]{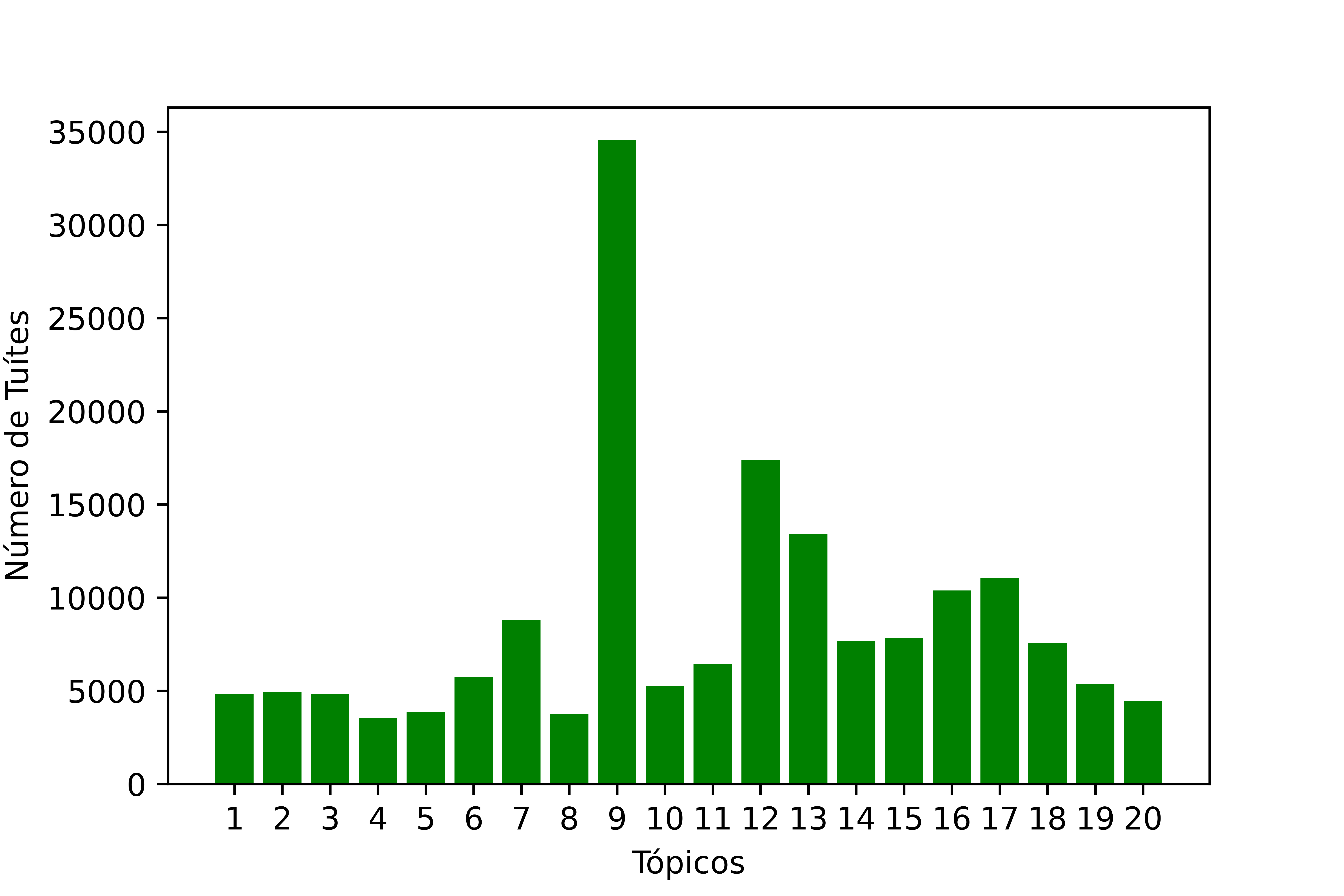}}
  \caption{Distribui\c{c}\~{a}o do n\'{u}mero de t\'{o}picos para cada m\^{e}s.}
  \label{fig:topicos-mes} 
\end{figure}

A Figura~\ref{fig:topicos-mes} apresenta a distribui\c{c}\~{a}o do n\'{u}mero de t\'{o}picos para cada m\^{e}s. N\'{o}s observamos que existe certos t\'{o}picos que s\~{a}o mais populares. Ao analisarmos o t\'{o}pico com maior n\'{u}mero de postagens em cada m\^{e}s, n\'{o}s percebemos que esse \'{e} um t\'{o}pico geral que agrupou termos relativos \`{a} pr\'{o}pria doen\c{c}a.

\subsection{Reconhecimento de Entidades Mencionadas (REM)}

Reconhecimento de Entidades Mencionadas (REM) \'{e} definida como uma tarefa n\~{a}o-trivial de automaticamente identificar e classificar certas men\c{c}\~{o}es a entidades em um dado texto~\cite{li2020survey}. Por exemplo, na senten\c{c}a ``Ch\'{a} de alho cura coronav\'{i}rus?'' postada por um dos usu\'{a}rios \'{e} poss\'{i}vel identificar uma d\'{u}vida sobre o uso de tratamentos alternativos para a cura da COVID-19. Neste caso, apesar de ch\'{a} de alho n\~{a}o ser um tipo de medicamento oficialmente reconhecido, o modelo REM \'{e} capaz de identificar que muitas pessoas consideram este tipo de ch\'{a} como um medicamento para a COVID-19.

Os m\'{e}todos que lidam com a tarefa de reconhecer entidades mencionadas s\~{a}o baseados principalmente em modelos de aprendizagem de m\'{a}quina~\cite{buyuktopacc2019evaluation,santos2019assessing,albared2019recent}. A extra\c{c}\~{a}o de entidades mencionadas no Twitter \'{e} uma tarefa ainda mais desafiadora~\cite{bontcheva2013twitie}. Primeiro, as postagens no Twitter s\~{a}o curtas (m\'{a}ximo de 280 caracteres) e, portanto, s\~{a}o mais dif\'{i}ceis de se interpretar quando comparadas com textos mais longos. Segundo, os textos curtos apresentam muitas varia\c{c}\~{o}es lingu\'{i}sticas e tendem a ser menos corretos em termos gramaticais. Por \'{u}ltimo, a maioria das pesquisas sobre ferramentas de processamento de linguagem natural s\~{a}o voltadas para o idioma ingl\^{e}s~\cite{santos2015named}.

Para esta tarefa, n\'{o}s usamos a ferramenta spaCy~\cite{spaCy:2020} que \'{e} baseada em modelos de redes neurais e permite a cria\c{c}\~{a}o de modelos pr\'{o}prios. O modelo de REM dispon\'{i}vel pelo spaCy em Portugu\^{e}s contempla apenas as seguintes entidades: Localiza\c{c}\~{a}o (LOC), Organiza\c{c}\~{a}o (ORG), Pessoa (PER) e Outros (MISC). Como o nosso objetivo neste trabalho \'{e} identificar entidades nas postagens relacionadas \`{a} COVID, n\'{o}s tivemos que criar um modelo pr\'{o}prio para reconhecer as seguintes entidades mencionadas: Rem\'{e}dios (DRUG), Doen\c{c}as (DIS), Pessoas (PER) e Organiza\c{c}\~{o}es (ORG). A raz\~{a}o para a escolha destas entidades \'{e} que as informa\c{c}\~{o}es relativas a essas categorias s\~{a}o essenciais durante uma crise de pandemia como a que estamos analisando. Apesar de existir bastante interesse sobre o desenvolvimento de modelos REM em Portugu\^{e}s~\cite{lopes2020comparing,fernandes2018applying,santos2019assessing}, nenhum dos modelos investigados poderia ser usado diretamente para reconhecer as entidades relacionadas \`{a} COVID-19 porque n\~{a}o foram treinados para reconhecer tais entidades.

N\'{o}s avaliamos o nosso modelo de REM sobre um conjunto de dados que foi manualmente anotado por n\'{o}s e ent\~{a}o mostramos que o modelo \'{e} efetivo na tarefa de identificar diversas categorias de entidades relacionadas \`{a} doen\c{c}a COVID-19. A discuss\~{a}o sobre a performance do modelo e sua aplica\c{c}\~{a}o ser\~{a}o discutidadas na pr\'{o}xima se\c{c}\~{a}o.


\section{Resultados e Discuss\~{o}es}
\label{sec:resultados}

\subsection{LDA Aplicado}

O nosso modelo final de LDA gerou 20 t\'{o}picos utilizando os valores padr\~{a}o de par\^{a}metros do modelo Gensim LDA MultiCore~\cite{rehurek_lrec}. A Tabela~\ref{tab:lda-resultados} apresenta tr\^{e}s t\'{o}picos escolhidos aleatoriamente por cada m\^{e}s do per\'{i}odo coletado de tu\'{i}tes (janeiro a abril de 2020). Para cada t\'{o}pico, n\'{o}s apresentamos os termos mais relevantes e tamb\'{e}m uma quest\~{a}o representativa que emprega tr\^{e}s desses termos por t\'{o}pico. Os termos est\~{a}o real\c{c}ados nas senten\c{c}as. Apesar de gerarmos os termos tamb\'{e}m como n-grams na fase de pr\'{e}-processamento, estes termos n\~{a}o tiveram influ\^{e}ncia na gera\c{c}\~{a}o dos t\'{o}picos. N\'{o}s avaliamos que isto ocorreu em raz\~{a}o do baixo n\'{u}mero de n-grams extra\'{i}dos. Todas as quest\~{o}es apresentadas foram realmente postadas por usu\'{a}rios do Twitter.

\begin{table*}[!h]
\centering
\caption{Principais t\'{o}picos gerados pelo modelo LDA para cada m\^{e}s.}
\scalebox{0.68}{
\begin{tabular}{@{\hspace{0.1cm}}c@{\hspace{0.1cm}}l@{\hspace{0.1cm}}l}
\hline
\textbf{T\'{o}pico} & \textbf{Termos}                                                                              & \textbf{Quest\~{a}o}                                                                                                       \\ \hline
1-Jan           & cerveja, pego, beber, ser, ano, transmitir, problema, beijar, vitimar                       & Se eu \hl{beber} \hl{cerveja} corona eu \hl{pego} corona v\'{i}rus? \\
2-Jan           & dengue, saber, lidar, epidemia, brasileiro, parecer, poder, significar      &  Brasil n\~{a}o \hl{sabe} \hl{lidar} com \hl{dengue}, vai saber lidar com corona virus?                                                                                                                      \\
3-Jan           & carnaval, chegar, risco, aqui, suspeito, entrar, imaginar, p\^{a}nico            &    Qual o \hl{risco} do coronav\'{i}rus \hl{chegar} no \hl{carnaval} no Brasil?                                                                                                                    \\ \hline
1-Fev           & carnaval, beijar, povo, ir, hoje, bom, tomar, continuar, familiar                       &   Ser\'{a} que nesse \hl{carnaval} o \hl{povo} vai \hl{beijar} menos com medo do coronav\'{i}rus?                                                                                                                    \\
2-Fev           & It\'{a}lia, causar, cancelar, estar, acontecer, virar, preocupar, tava, evitar         & Vai \hl{cancelar} viagem para \hl{Italia} por \hl{causa} do coronav\'{i}rus?                                                                                                                       \\
3-Fev           & pandemia, epidemia, global, tamb\'{e}m, verdade, tempo, mundial, real        &   \hl{Epidemia} \hl{global} n\~{a}o seria \hl{pandemia}?                                                                                                                     \\ \hline
1-Mar           & gente, morrer, causar, ver, ter, sair, assim, ajudar, pegar                                                                                             &  \hl{Gente} e se eu \hl{morrer} por \hl{causa} desse corona v\'{i}rus?                                                                                                                      \\
2-Mar           & sobreviver, ser\'{a}, pessoa, algu\'{e}m, voltar, ruir, explicar, necessidade                                                                                             & Como \hl{ser\'a} que essas \hl{pessoas} v\~{a}o \hl{sobreviver}?                                                                                                                       \\
3-Mar           & pandemia, fazer, causa, testar, pleno, epidemia, gripezinha, atacar                                                                                             &  O que \hl{fazer} para minimizar a crise \hl{causada} pela \hl{pandemia} do coronav\'{i}rus?                                                                                                                      \\ \hline
1-Abr           & social, isolamento, querer, entender, partir, medir, campanha, decidir                                                                                             & Como passar por um \hl{isolamento} \hl{social} sem \hl{querer} se matar?                                                                                                                       \\
2-Abr           &    pandemia, presidente, ministro, morte, passar, tirar, explicar, apenas                                                                                          &    O \hl{presidente} demitindo o \hl{ministro} da sa\'{u}de no meio de uma \hl{pandemia}?                                                                                                                    \\
3-Abr           &    curar, tempo, achar, ter, ir, bem, hoje, liberar                                                                                          & Voc\^{e} \hl{acha} que \hl{tempo} \'{e} \hl{cura} pra v\'{i}rus?    \\ \hline                                                                                                                  
\end{tabular}
}
\label{tab:lda-resultados}
\end{table*}

A not\'{i}cia de pessoas infectadas com a COVID-19 na China foi recebida com um certo grau de ironia e incredulidade nas perguntas postadas pelos usu\'{a}rios. O T\'{o}pico 1-Jan exibido na Tabela~\ref{tab:lda-resultados} mostra uma pergunta ir\^{o}nica onde o usu\'{a}rio do Twitter faz uma brincadeira relacionando o nome da doen\c{c}a com o nome de uma famosa marca de cerveja. J\'{a} no T\'{o}pico 3-Jan, observa-se que a preocupa\c{c}\~{a}o para muitos dos usu\'{a}rios era saber se a doen\c{c}a poderia atrapalhar a realiza\c{c}\~{a}o do carnaval no Brasil.

N\'{o}s observarmos uma certa mudan\c{c}a da percep\c{c}\~{a}o dos usu\'{a}rios em rela\c{c}\~{a}o \`{a} doen\c{c}a no m\^{e}s seguinte. Apesar das pessoas ainda questionarem sobre a realiza\c{c}\~{a}o ou n\~{a}o do carnaval no Brasil, conforme descrito no t\'{o}pico 1-Fev, as pessoas come\c{c}aram a fazer perguntas para entender melhor a terminologia da doen\c{c}a. \'{E} o que est\'{a} representado na pergunta do t\'{o}pico 3-Fev, onde o usu\'{a}rio questiona a diferen\c{c}a entre pandemia e endemia. Este foi um assunto bastante comentado pelas reportagens publicadas por meios de comunica\c{c}\~{a}o no Pa\'{i}s. Ainda neste m\^{e}s de fevereiro, um questionamento comum entre os usu\'{a}rios foi saber o alcance da doen\c{c}a em pa\'{i}ses afastados da China, como foi o caso da It\'{a}lia (t\'{o}pico 2-Fev).

No m\^{e}s de mar\c{c}o ocorreu uma n\'{i}tida mudan\c{c}a dos questionamentos dos usu\'{a}rios do Twitter em face \`{a} doen\c{c}a. Termos como ``morrer'' e ``sobreviver'' foram bastante utilizados pelos usu\'{a}rios, conforme pode ser observado nos t\'{o}picos 1-Mar e 2-Mar. Isto \'{e} devido, especialmente, pelo fato da morte do primeiro brasileiro por COVID-19. Um outro t\'{o}pico que apareceu com frequ\^{e}ncia no m\^{e}s de mar\c{c}o foi a preocupa\c{c}\~{a}o das pessoas com as a\c{c}\~{o}es governamentais para minimizar o efeito da pandemia na sa\'{u}de p\'{u}blica e tamb\'{e}m na economia do Pa\'{i}s. Isto pode visualizado no t\'{o}pico 3-Mar.

No m\^{e}s de abril, o principal foco dos usu\'{a}rios foi nas a\c{c}\~{o}es de governantes para lidar com o alastramento da doen\c{c}a. Conforme observamos no t\'{o}pico 1-Abr, muitos usu\'{a}rios questionaram a efici\^{e}ncia do isolamento social e \textit{lockdown} para minimizar a propaga\c{c}\~{a}o da doen\c{c}a. Um outro t\'{o}pico bastante ativo nos questionamentos dos usu\'{a}rios foram as discuss\~{o}es pol\'{i}ticas. \'{E} o que podemos observar no t\'{o}pico 2-Abr, onde um usu\'{a}rio questiona a decis\~{a}o do presidente do Brasil de demitir o Ministro da Sa\'{u}de durante o per\'{i}odo de pandemia. Finalmente, uma outra d\'{u}vida bastante comum entre os usu\'{a}rios neste m\^{e}s de abril foi a efic\'{a}cia dos rem\'{e}dios e outros tratamentos para combater a doen\c{c}a. O t\'{o}pico 3-Abr ilustra a preocupa\c{c}\~{a}o dos usu\'{a}rios com a cura da doen\c{c}a.

\subsection{Modelo REM Aplicado}

Nesta se\c{c}\~{a}o, n\'{o}s discutimos a cria\c{c}\~{a}o e a performance do modelo REM proposto e tamb\'{e}m a aplica\c{c}\~{a}o do modelo no conjunto de dados coletados. Para o modelo REM proposto, n\'{o}s consideramos quatro entidades: Doen\c{c}a (DIS), Medicamento (DRUG), Organiza\c{c}\~{a}o (ORG) e Pessoa (PER). Para cada tipo de entidade mencionada, n\'{o}s usamos as m\'{e}tricas de \textit{Precision} (P), \textit{Recall} (R) e \textit{F-Measure} (F1). \textit{Precision} corresponde ao percentual correto de men\c{c}\~{o}es a entidades, \textit{recall} corresponde ao percentual do total de entidades que foram corretamente reconhecidas pelo modelo, enquanto \textit{F-Measure} (F1) \'{e} a m\'{e}dia harm\^{o}nica entre precis\~{a}o e revoca\c{c}\~{a}o. Estas m\'{e}tricas s\~{a}o comumente utilizadas para avaliar a performance de modelos de REM~\cite{li2020survey,jung2015ln,liu2011recognizing,derczynski2015analysis}.

\begin{figure}[!h]
\centering
\includegraphics[width=0.45\textwidth]{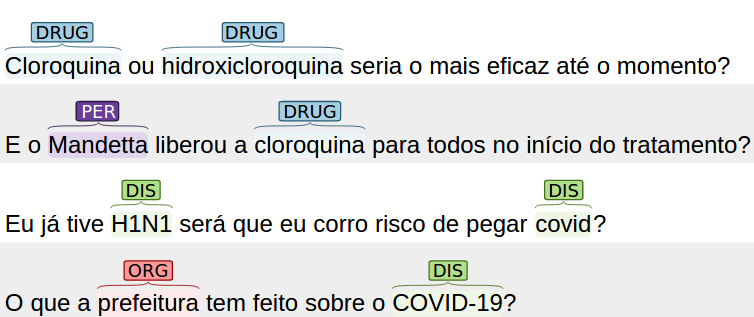}
\caption{Exemplo de entidades anotadas.}
\label{fig:webanno}
\end{figure}

Com o objetivo de treinar o novo modelo, n\'{o}s fizemos a anota\c{c}\~{a}o dos dados de treino. Para esta tarefa, n\'{o}s usamos a ferramenta WebAnno~\cite{de2016web}. A Figura~\ref{fig:webanno} mostra uma captura de tela da ferramenta e alguns exemplos de anota\c{c}\~{o}es das entidades. 

N\'{o}s fizemos a anota\c{c}\~{a}o em 2.000 das perguntas coletadas do Twitter selecionadas aleatoriamente. Este \textit{dataset} anotado foi dividido em 80\% para treino e 20\% para testes. Tabela~\ref{tab:resultados-rem} apresenta os resultados alcan\c{c}ados pelo modelo REM para cada categoria. 

\begin{table}[h]
\centering
\caption{Resultados alcan\c{c}ados pelo modelo REM desenvolvido.}
\begin{tabular}{cccc}
\hline
\textbf{Categoria} & \textbf{Precision} & Recall & \textbf{F1} \\ \hline
Doen\c{c}a             & 98,97             & 96,50              & 97,72       \\ 
Medicamentos       & 87,80              & 88,26              & 88,04       \\ 
Organiza\c{c}\~{a}o        & 93,40             & 92,92              & 93,16       \\ 
Pessoa             & 95,00                & 96,44              & 95,71       \\ \hline
\end{tabular}
\label{tab:resultados-rem}
\end{table}

De acordo com os resultados obtidos, a m\'{e}dia de F1 entre as quatro entidades foi 93,65. O desempenho mais baixo do modelo foi na categoria de Medicamentos. N\'{o}s acreditamos que isso ocorreu porque o modelo identificou diversas subst\^{a}ncias que t\^{e}m efeito sobre as pessoas, mas que n\~{a}o foram anotadas como medicamento. \'{E} o caso, por exemplo, dos v\'{a}rios tipos de ch\'{a} que s\~{a}o mencionados pelas pessoas como alternativas para o tratamento da doen\c{c}a.

\begin{figure}[!h]
    \centering
  \subfloat[Doen\c{c}a (DIS)\label{1a}]{%
       \includegraphics[width=0.45\linewidth]{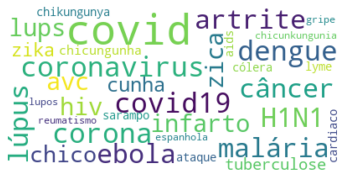}}
    \hfill
  \subfloat[Medicamento (DRUG)\label{1b}]{%
        \includegraphics[width=0.45\linewidth]{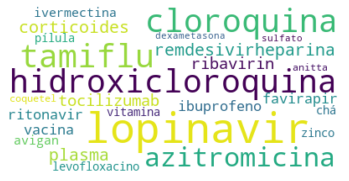}}
    \\
  \subfloat[Organiza\c{c}\~{a}o (ORG)\label{1c}]{%
        \includegraphics[width=0.45\linewidth]{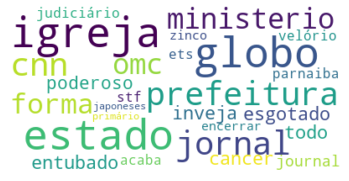}}
    \hfill 
  \subfloat[Pessoa (PER)\label{1d}]{%
        \includegraphics[width=0.45\linewidth]{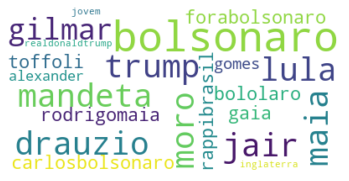}}
  \caption{(a) Doen\c{c}a. (b) Medicamento. (c) Pessoas. (d) Organiza\c{c}\~{a}o.}
  \label{fig:rem} 
\end{figure}

Na Figura~\ref{fig:rem} est\~{a}o representadas as men\c{c}\~{o}es \`{a}s entidades encontradas para cada uma das categorias consideradas neste trabalho usando o modelo discutido acima. Denominada de nuvem de palavras~\cite{heimerl2014word}, esta informa\c{c}\~{a}o fornece um meio simples de comunicar visualmente as palavras mais frequentemente usadas em cada categoria. Assim, pode-ser perceber que os termos mais frequentes est\~{a}o alinhados com as d\'{u}vidas dos usu\'{a}rios do Twitter.

Na Figura~\ref{fig:rem}(a), n\'{o}s apresentamos as men\c{c}\~{o}es mais frequentes em perguntas relacionadas \`{a} doen\c{c}a. Primeiramente, \'{e} poss\'{i}vel observar que a doen\c{c}a \'{e} nomeada de diversas maneiras. COVID, COVID-19 e corona s\~{a}o as formas mais comuns de nomear a doen\c{c}a. Al\'{e}m da COVID e suas diversas denomina\c{c}\~{o}es, as pessoas tamb\'{e}m fizeram muitos questionamentos sobre outras doen\c{c}as. Comumente, as perguntas sobre as demais doen\c{c}as fazem uma rela\c{c}\~{a}o com a COVID-19. Por exemplo, um usu\'{a}rio perguntou \textit{``Algu\'{e}m sabe me dizer a diferen\c{c}a dos sintomas de H1N1 para a COVID19?''}. A compara\c{c}\~{a}o entre os sintomas da COVID-19 e de outras doen\c{c}as foi o tipo de pergunta mais comum entre os usu\'{a}rios.

Na Figura~\ref{fig:rem}(b), n\'{o}s apresentamos as men\c{c}\~{o}es mais frequentes em perguntas relacionadas aos principais medicamentos usados para o tratamento da doen\c{c}a. D\'{u}vidas sobre quais medicamentos usar e a efic\'{a}cia desses medicamentos no tratamento da doen\c{c}a foram as d\'{u}vidas mais comuns que encontramos durante a an\'{a}lise das postagens. Cloroquina, hidroxicloroquina e azitromicina foram os medicamentos mais mencionados nas perguntas dos usu\'{a}rios. 

Interessantemente, o termo ch\'{a} foi identificado pelo nosso modelo de REM como uma men\c{c}\~{a}o a medicamentos. Ao analisarmos o contexto do uso da men\c{c}\~{a}o de ch\'{a} nas quest\~{o}es dos usu\'{a}rios, n\'{o}s identificamos a d\'{u}vida comum se alguns tipos de ch\'{a} poderiam ser usados no combate \`{a} COVID-19. A seguir, apresentamos alguns exemplos reais de perguntas que mencionaram o termo ch\'{a}: a) \textit{`` ch\'{a} de anis estrelado ajuda na preven\c{c}\~{a}o ou aumenta a umidade (sic) por conta do tamiflu?''}; b) \textit{Para combater o novo coronav\'{i}rus, m\'{e}dicos recomendaram tomar ch\'{a} de erva-doce por ele conter a subst\^{a}ncia do Tamiflu?}; c) \textit{``Ch\'{a} de boldo para coronav\'{i}rus?''}.

A identifica\c{c}\~{a}o das d\'{u}vidas mais comuns sobre o uso dos medicamentos para os usu\'{a}rios ajudaria aos agentes p\'{u}blicos no enfretamento da doen\c{c}a. Por exemplo, os governantes poderiam usar esse conhecimento para divulgar campanhas esclarecendo as principais d\'{u}vidas sobre o uso de determinados medicamentos. Uma campanha direta informando, por exemplo, o que \'{e} fato ou mito.

Na Figura~\ref{fig:rem}(c), n\'{o}s apresentamos as men\c{c}\~{o}es mais frequentes em perguntas relacionadas \`{a}s organiza\c{c}\~{o}es diante da pandemia de COVID-19. Estado, Minist\'{e}rio (da Sa\'{u}de), Globo, Congresso e STF foram as organiza\c{c}\~{o}es mais comumente mencionadas nas perguntas postadas pelos usu\'{a}rios. A men\c{c}\~{a}o a Estado costuma aparecer nas perguntas se referindo ao Governo Federal ou \`{a} alguma das Unidades da Federa\c{c}\~{a}o, como S\~{a}o Paulo, Rio de Janeiro ou Amazonas. Ao analisarmos o contexto em que essas men\c{c}\~{o}es aparecem, n\'{o}s observamos que as perguntas costumam ter uma cr\'{i}tica a essas organiza\c{c}\~{o}es. A emissora Rede Globo \'{e} a terceira organiza\c{c}\~{a}o mais mencionada nas perguntas. A seguir, apresentamos exemplos de perguntas que ilustram essa nossa observa\c{c}\~{a}o: a) \textit{``quando v\~{a}o avisar a gente q (sic) o coronav\'{i}rus \'{e} uma grande pegadinha da globo e do pt?''}; b) \textit{``Quanto voc\^{e} e a Globo v\~{a}o doar para o combate ao corona v\'{i}rus?''}; c) \textit{``Uma PANDEMIA acontecendo e a Globo vai se preocupar com uma cantora fumando maconha?''}. A identifica\c{c}\~{a}o deste tipo de questionamento pode ser usado pelas organiza\c{c}\~{o}es para compreenderem a vis\~{a}o que uma parte representativa de pessoas t\^{e}m sobre a conduta dessas organiza\c{c}\~{o}es perante a pandemia.

Na Figura~\ref{fig:rem}(d), n\'{o}s apresentamos as men\c{c}\~{o}es mais frequentes \`{a}s pessoas em perguntas relacionadas a pandemia de COVID-19. Os nomes que mais apareceram nessa an\'{a}lise foram de pol\'{i}ticos, com ou sem mandato, e pessoas p\'{u}blicas que constatemente aparecem na grande m\'{i}dia. Este \'{e} o caso, por exemplo, do m\'{e}dico Dr\'{a}uzio Varela da Rede Globo que ficou em bastante evid\^{e}ncia ap\'{o}s algumas declara\c{c}\~{o}es pol\^{e}micas na TV~\cite{BBC-Drauzio:2020}.

Assim como ocorreu na aplica\c{c}\~{a}o do LDA para gera\c{c}\~{a}o de t\'{o}picos, os termos representados como n-grams n\~{a}o apareceram destacados na nuvem de palavras.

\section{Conclus\~{o}es e Trabalhos Futuros}
\label{sec:conclusoes}

N\'{o}s iniciamos esta se\c{c}\~{a}o discutindo algumas limita\c{c}\~{o}es atribu\'{i}das a este estudo. Os dados foram obtidos atrav\'{e}s de um conjunto pr\'{e}-definido de palavras-chaves. Portanto, \'{e} poss\'{i}vel que alguns usu\'{a}rios tenham publicado mensagens sobre a pandemia de COVID-19 usando um conjunto de termos diferentes das palavras-chaves usadas na coleta e, consequentemente, estas mensagens n\~{a}o foram coletadas. Outra limita\c{c}\~{a}o \'{e} que o Twitter n\~{a}o divulga dados sobre o perfil de seus usu\'{a}rios, tais como idade, sexo ou classe social. Assim, n\~{a}o foi poss\'{i}vel realizar uma an\'{a}lise estratificada dos usu\'{a}rios e a amostragem analisada pode n\~{a}o representar toda a popula\c{c}\~{a}o brasileira. Al\'{e}m disso, uma limita\c{c}\~{a}o \'{e} que focamos exclusivamente nos dados de redes sociais e, portanto, nossas conclus\~{o}es n\~{a}o podem ser generaliz\'{a}veis para outras plataformas de m\'{i}dias sociais ou outros meios de comunica\c{c}\~{a}o, tais como TV ou r\'{a}dio. Finalmente, apesar do Twitter ser uma plataforma bastante popular no Brasil, esta n\~{a}o \'{e} utilizada por toda a popula\c{c}\~{a}o. Assim, recomenda-se cautela antes de assumir a generaliza\c{c}\~{a}o dos resultados.

A pandemia do COVID-19 vem ceifando a vida de milh\~{o}es de pessoas no mundo. Atualmente, muitas pessoas fazem uso das m\'{i}dias sociais como o Twitter para expressar diversos tipos de questionamentos sobre a doen\c{c}a. A compreens\~{a}o das d\'{u}vidas comuns dos usu\'{a}rios dessas redes sociais pode ser um ponto de partida para projetar mensagens estrat\'{e}gicas para campanhas de sa\'{u}de e estabelecer um sistema de comunica\c{c}\~{a}o eficaz durante a pandemia para um melhor enfrentamento \`{a} doen\c{c}a.

Como trabalhos futuros, n\'{o}s pretendemos investigar a aplica\c{c}\~{a}o dos m\'{e}todos desenvolvidos neste trabalho em outras fontes de m\'{i}dias sociais, tais como, Instagram, Facebook e YouTube. Al\'{e}m disso, n\'{o}s pretendemos estender o estudo para incluir dados de per\'{i}odos mais longos, mesmo ap\'{o}s o fim da pandemia. O objetivo \'{e} entender a manifesta\c{c}\~{a}o das pessoas sobre o surto da doen\c{c}a.

\bibliographystyle{plain}
\bibliography{references}        

\begin{thebibliography}{10}

\bibitem{googletrends:2020}
{\em Google Trends}, acessado em 26 de maio de 2020.

\bibitem{twitterscraper:2020}
{\em Twitterscraper}, acessado em 26 de maio de 2020.

\bibitem{abd2020top}
Alaa Abd-Alrazaq, Dari Alhuwail, Mowafa Househ, Mounir Hamdi, and Zubair Shah.
\newblock Top concerns of tweeters during the covid-19 pandemic: infoveillance
  study.
\newblock {\em Journal of medical Internet research}, 22(4):e19016, 2020.
\newblock DOI:\url{10.2196/19016}.

\bibitem{albared2019recent}
Mohammed Albared, Marc~Gallofr{\'e} Oca{\~n}a, Abdullah Ghareb, and Tareq
  Al-Moslmi.
\newblock Recent progress of named entity recognition over the most popular
  datasets.
\newblock In {\em 2019 First International Conference of Intelligent Computing
  and Engineering (ICOICE)}, pages 1--9. IEEE, 2019.
\newblock DOI:\url{10.1109/ICOICE48418.2019.9035170}.

\bibitem{blei2003latent}
David~M Blei, Andrew~Y Ng, and Michael~I Jordan.
\newblock Latent dirichlet allocation.
\newblock {\em Journal of machine Learning research}, 3(Jan):993--1022, 2003.
\newblock DOI:\url{10.1162/jmlr.2003.3.4-5.993}.

\bibitem{bontcheva2013twitie}
Kalina Bontcheva, Leon Derczynski, Adam Funk, Mark~A Greenwood, Diana Maynard,
  and Niraj Aswani.
\newblock Twitie: An open-source information extraction pipeline for microblog
  text.
\newblock In {\em Proceedings of the International Conference Recent Advances
  in Natural Language Processing RANLP 2013}, pages 83--90, 2013.
\newblock DOI:\url{10.6084/M9.FIGSHARE.1003767.V2}.

\bibitem{BBC-Drauzio:2020}
BBC Brasil.
\newblock {\em Drauzio Varella prev\^{e} trag\'{e}dia nacional por
  coronav\'{i}rus: Brasil vai pagar o pre\c{c}o da desigualdade}, 2020
  (accessed May 10, 2020).

\bibitem{BBC:2020}
BBC Brasil.
\newblock {\em Coronav\'{i}rus: SP confirma novas mortes e n\'{u}mero de
  \'{o}bitos sobe para 18 no Brasil}, 2020 (accessed May 20, 2020).

\bibitem{buyuktopacc2019evaluation}
Onur B{\"u}y{\"u}ktopa{\c{c}} and Tankut Acarman.
\newblock Evaluation of cosine similarity feature for named entity recognition
  on tweets.
\newblock In {\em International Conference on Man--Machine Interactions}, pages
  125--135. Springer, 2019.
\newblock DOI:\url{10.1007/978-3-030-31964-9_12}.

\bibitem{de2016web}
Richard~Eckart de~Castilho, Eva Mujdricza-Maydt, Seid~Muhie Yimam, Silvana
  Hartmann, Iryna Gurevych, Anette Frank, and Chris Biemann.
\newblock A web-based tool for the integrated annotation of semantic and
  syntactic structures.
\newblock In {\em Proceedings of the Workshop on Language Technology Resources
  and Tools for Digital Humanities (LT4DH)}, pages 76--84, 2016.

\bibitem{derczynski2015analysis}
Leon Derczynski, Diana Maynard, Giuseppe Rizzo, Marieke Van~Erp, Genevieve
  Gorrell, Rapha{\"e}l Troncy, Johann Petrak, and Kalina Bontcheva.
\newblock Analysis of named entity recognition and linking for tweets.
\newblock {\em Information Processing \& Management}, 51(2):32--49, 2015.
\newblock DOI:\url{10.1016/j.ipm.2014.10.006}.

\bibitem{du2019twitter}
Hanxiang Du, Long Nguyen, Zhou Yang, Hashim Abu-Gellban, Xingyu Zhou, Wanli
  Xing, Guofeng Cao, and Fang Jin.
\newblock Twitter vs news: Concern analysis of the 2018 california wildfire
  event.
\newblock In {\em 2019 IEEE 43rd Annual Computer Software and Applications
  Conference (COMPSAC)}, volume~2, pages 207--212. IEEE, 2019.
\newblock DOI:\url{10.1109/COMPSAC.2019.10208}.

\bibitem{fernandes2018applying}
Ivo Fernandes, Henrique~Lopes Cardoso, and Eugenio Oliveira.
\newblock Applying deep neural networks to named entity recognition in
  portuguese texts.
\newblock In {\em 2018 Fifth International Conference on Social Networks
  Analysis, Management and Security (SNAMS)}, pages 284--289. IEEE, 2018.
\newblock DOI:\url{10.1109/SNAMS.2018.8554782}.

\bibitem{gopi2020classification}
Arepalli~Peda Gopi, R~Naga~Sravana Jyothi, V~Lakshman Narayana, and K~Satya
  Sandeep.
\newblock Classification of tweets data based on polarity using improved rbf
  kernel of svm.
\newblock {\em International Journal of Information Technology}, pages 1--16,
  2020.
\newblock DOI:\url{10.1007/s41870-019-00409-4}.

\bibitem{heimerl2014word}
Florian Heimerl, Steffen Lohmann, Simon Lange, and Thomas Ertl.
\newblock Word cloud explorer: Text analytics based on word clouds.
\newblock In {\em 47th Hawaii International Conference on System Sciences},
  pages 1833--1842. IEEE, 2014.
\newblock DOI:\url{10.1109/HICSS.2014.231}.

\bibitem{spaCy:2020}
Matthew Honnibal.
\newblock {\em spaCy}, acessado em 26 de maio de 2020.

\bibitem{jelodar2019latent}
Hamed Jelodar, Yongli Wang, Chi Yuan, Xia Feng, Xiahui Jiang, Yanchao Li, and
  Liang Zhao.
\newblock Latent dirichlet allocation (lda) and topic modeling: models,
  applications, a survey.
\newblock {\em Multimedia Tools and Applications}, 78(11):15169--15211, 2019.
\newblock DOI:\url{10.1145/1027154.1027165}.

\bibitem{jung2015ln}
YoungHoon Jung, Karl Stratos, and Luca~P Carloni.
\newblock Ln-annote: An alternative approach to information extraction from
  emails using locally-customized named-entity recognition.
\newblock In {\em Proceedings of the 24th International Conference on World
  Wide Web}, pages 538--548, 2015.
\newblock DOI:\url{10.1145/2736277.2741633}.

\bibitem{kim2016topic}
Erin Hea-Jin Kim, Yoo~Kyung Jeong, Yuyoung Kim, Keun~Young Kang, and Min Song.
\newblock Topic-based content and sentiment analysis of ebola virus on twitter
  and in the news.
\newblock {\em Journal of Information Science}, 42(6):763--781, 2016.
\newblock DOI:\url{10.1177/0165551515608733}.

\bibitem{li2011question}
Baichuan Li, Xiance Si, Michael~R Lyu, Irwin King, and Edward~Y Chang.
\newblock Question identification on twitter.
\newblock In {\em Proceedings of the 20th ACM international conference on
  Information and knowledge management}, pages 2477--2480, 2011.
\newblock DOI:\url{10.1145/2063576.2063996}.

\bibitem{li2020survey}
Jing Li, Aixin Sun, Jianglei Han, and Chenliang Li.
\newblock A survey on deep learning for named entity recognition.
\newblock {\em IEEE Transactions on Knowledge and Data Engineering}, 2020.
\newblock DOI:\url{10.1109/TKDE.2020.2981314}.

\bibitem{li2020characterizing}
Lifang Li, Qingpeng Zhang, Xiao Wang, Jun Zhang, Tao Wang, Tian-Lu Gao, Wei
  Duan, Kelvin Kam-fai Tsoi, and Fei-Yue Wang.
\newblock Characterizing the propagation of situational information in social
  media during covid-19 epidemic: A case study on weibo.
\newblock {\em IEEE Transactions on Computational Social Systems},
  7(2):556--562, 2020.
\newblock DOI:\url{10.1109/TCSS.2020.2980007}.

\bibitem{liu2020health}
Qian Liu, Zequan Zheng, Jiabin Zheng, Qiuyi Chen, Guan Liu, Sihan Chen, Bojia
  Chu, Hongyu Zhu, Babatunde Akinwunmi, Jian Huang, et~al.
\newblock Health communication through news media during the early stage of the
  covid-19 outbreak in china: Digital topic modeling approach.
\newblock {\em Journal of medical Internet research}, 22(4):e19118, 2020.
\newblock DOI:\url{10.2196/19118}.

\bibitem{liu2011recognizing}
Xiaohua Liu, Shaodian Zhang, Furu Wei, and Ming Zhou.
\newblock Recognizing named entities in tweets.
\newblock In {\em Proceedings of the 49th Annual Meeting of the Association for
  Computational Linguistics: Human Language Technologies}, pages 359--367,
  2011.

\bibitem{lopes2020comparing}
F{\'a}bio Lopes, C{\'e}sar Teixeira, and Hugo~Gon{\c{c}}alo Oliveira.
\newblock Comparing different methods for named entity recognition in
  portuguese neurology text.
\newblock {\em Journal of Medical Systems}, 44(4):1--20, 2020.
\newblock DOI:\url{10.1007/s10916-020-1542-8}.

\bibitem{malta2020coronavirus}
Monica Malta, Anne~W Rimoin, and Steffanie~A Strathdee.
\newblock The coronavirus 2019-ncov epidemic: Is hindsight 20/20?
\newblock {\em EClinicalMedicine}, 20, 2020.
\newblock DOI:\url{10.1016/j.eclinm.2020.100289}.

\bibitem{ordun2020exploratory}
Catherine Ordun, Sanjay Purushotham, and Edward Raff.
\newblock Exploratory analysis of covid-19 tweets using topic modeling, umap,
  and digraphs.
\newblock {\em arXiv preprint arXiv:2005.03082}, 2020.

\bibitem{world2020coronavirus}
World~Health Organization et~al.
\newblock Coronavirus disease 2019 (covid-19): situation report, 126.
\newblock 2020.

\bibitem{paul2011twitter}
Sharoda~A Paul, Lichan Hong, and Ed~H Chi.
\newblock Is twitter a good place for asking questions? a characterization
  study.
\newblock In {\em Fifth International AAAI Conference on Weblogs and Social
  Media}, 2011.

\bibitem{rehurek_lrec}
Radim {\v R}eh{\r u}{\v r}ek and Petr Sojka.
\newblock {Software Framework for Topic Modelling with Large Corpora}.
\newblock In {\em {Proceedings of the LREC 2010 Workshop on New Challenges for
  NLP Frameworks}}, pages 45--50, Valletta, Malta, May 2010. ELRA.

\bibitem{roder2015exploring}
Michael R{\"o}der, Andreas Both, and Alexander Hinneburg.
\newblock Exploring the space of topic coherence measures.
\newblock In {\em Proceedings of the eighth ACM international conference on Web
  search and data mining}, pages 399--408, 2015.
\newblock DOI:\url{10.1145/2684822.2685324}.

\bibitem{santos2015named}
Joao Tiago~Luis Santos, Ivo~Miguel Anastacio, and Bruno~Emanuel Martins.
\newblock Named entity disambiguation over texts written in the portuguese or
  spanish languages.
\newblock {\em IEEE Latin America Transactions}, 13(3):856--862, 2015.
\newblock DOI:\url{10.1109/TLA.2015.7069115}.

\bibitem{santos2019assessing}
Joaquim Santos, Bernardo Consoli, Cicero dos Santos, Juliano Terra, Sandra
  Collonini, and Renata Vieira.
\newblock Assessing the impact of contextual embeddings for portuguese named
  entity recognition.
\newblock In {\em 2019 8th Brazilian Conference on Intelligent Systems
  (BRACIS)}, pages 437--442. IEEE, 2019.
\newblock DOI:\url{10.1109/BRACIS.2019.00083}.

\bibitem{sinnenberg2017twitter}
Lauren Sinnenberg, Alison~M Buttenheim, Kevin Padrez, Christina Mancheno, Lyle
  Ungar, and Raina~M Merchant.
\newblock Twitter as a tool for health research: a systematic review.
\newblock {\em American journal of public health}, 107(1):e1--e8, 2017.
\newblock DOI:\url{10.2105/AJPH.2016.303512}.

\bibitem{soulier2016answering}
Laure Soulier, Lynda Tamine, and Gia-Hung Nguyen.
\newblock Answering twitter questions: a model for recommending answerers
  through social collaboration.
\newblock In {\em Proceedings of the 25th ACM International on Conference on
  Information and Knowledge Management}, pages 267--276, 2016.
\newblock DOI:\url{10.1145/2983323.2983771}.

\bibitem{syed2017full}
Shaheen Syed and Marco Spruit.
\newblock Full-text or abstract? examining topic coherence scores using latent
  dirichlet allocation.
\newblock In {\em 2017 IEEE International conference on data science and
  advanced analytics (DSAA)}, pages 165--174. IEEE, 2017.
\newblock DOI:\url{10.1109/DSAA.2017.61}.

\bibitem{velavan2020covid}
Thirumalaisamy~P Velavan and Christian~G Meyer.
\newblock The covid-19 epidemic.
\newblock {\em Tropical medicine \& international health}, 25(3):278, 2020.

\bibitem{zahra2017geographic}
Kiran Zahra, Frank~O Ostermann, and Ross~S Purves.
\newblock Geographic variability of twitter usage characteristics during
  disaster events.
\newblock {\em Geo-spatial information science}, 20(3):231--240, 2017.
\newblock DOI:\url{10.1080/10095020.2017.1371903}.

\bibitem{zhao2013questions}
Zhe Zhao and Qiaozhu Mei.
\newblock Questions about questions: An empirical analysis of information needs
  on twitter.
\newblock In {\em Proceedings of the 22nd international conference on World
  Wide Web}, pages 1545--1556, 2013.
\newblock DOI:\url{10.1145/2488388.2488523}.

\end{thebibliography}

\end{document}